\documentclass[11pt,a4paper]{article}
\usepackage{moriond,epsfig,amsmath,floatflt}

\def\Journal#1#2#3#4{{#1} {\bf #2}, #3 (#4)}

\def\PRL{\em Phys.\@ Rev.\@ Lett.\@ }
\def\PRD{{\em Phys. Rev.} D}
\def\ZPC{{\em Z. Phys.} C}
\def\EPC{{\em Eur.\@ Phys.\@ J.\@ }C}

\def\ea{{\em et al.}}

\def\be{\begin{equation}}
\def\ee{\end{equation}}
\def\bea{\begin{eqnarray}}
\def\eea{\end{eqnarray}}

\newcommand{\pom}{I\!\!P}
\newcommand{\xpom}{x_{\pom}}
\newcommand{\zpom}{z_{\pom}}

\begin{document}


\vspace*{4cm}
\title{
\mathversion{bold}
DIFFRACTION IN $ep$ COLLISIONS
\mathversion{normal}}

\author{PIERRE VAN MECHELEN \footnote{Postdoctoral fellow of the Fund
for Scientific Research - Flanders (Belgium);
Pierre.VanMechelen@ua.ac.be}}

\address{Universiteit Antwerpen}

\maketitle\abstracts{Recent measurements of the diffractive
deep-inelastic cross section are used to extract diffractive parton
densities of the proton.  These are subsequently applied in models to
predict the production of jets and open charm in the final state.  A
rapidity gap suppression factor for dijet production in diffractive photoproduction
relative to diffractive deep-inelastic scattering is obtained using a model-independent comparison.}

\section{Introduction}

The measurement of diffractive processes in deep-inelastic scattering
(DIS), characterized by the presence of a large rapidity gap in the
final state, has been used to extract so-called diffractive parton
densities of the proton.  According to~\cite{bib:collins}, this is a
valid procedure, since the QCD hard scattering factorization theorem
should also hold for diffractive $ep$ collisions, yielding parton
density functions that are conditional on the presence of a proton with
fixed four-momentum in the final state.  It has also been pointed out,
however, that using these conditional parton densities to predict jet
rates in $p\overline{p}$ collisions, leads to an overestimation of the
cross section by an order of magnitude~\cite{bib:cdfjets}.  This
breaking of factorization between $ep$ and $p\overline{p}$ interactions
is generally attributed to rescattering processes that are present in
$p\overline{p}$ but not in $ep$ collisions, and has been succesfully
parameterized as a ``rapidity gap survival
probability''~\cite{bib:kaidalov}.

In $ep$ collisions mediated by quasi-real photons, one distinguishes
(in leading order (LO) QCD) direct processes, where the exchanged photon
interacts as a whole, from resolved processes, where the photon is
treated as a source of partons, one of which produces a hard scattering
with the proton, leaving a photon remnant behind.  This last class of
processes, where the photon has a hadron-like component, is reminiscent
of $p\overline{p}$ collisions and thus naturally leads to the question
whether rapidity gap suppression is also observed here.

This paper first discusses new measurements of the inclusive
diffractive deep-inelastic cross section and the extraction of
diffractive parton densities.  These results are subsequently used to
interpret results on dijet and open charm production in diffractive
DIS and to look for rapidity gap suppression in diffractive
photoproduction of dijet events.

\section{Inclusive diffractive deep-inelastic scattering}

\subsection{Cross section measurements}

In addition to the usual DIS ki\-nematic variables, the photon
virtuality $Q^2$ and the Bj\"orken scaling variables $x$ and $y$, one
introduces in the case of diffractive DIS the variables $\xpom$ and
$\beta$, respectively defined as the longitudinal momentum fraction of
the proton carried by the colourless exchange causing the rapidity gap, and
the longitudinal momentum fraction of the colourless exchange carried
by the struck quark.  $\beta$ has an analogous interpretation in the
$\gamma \pom$ \footnote{$\pom$ is a generic label used for the
colourless exchange, which in some models is identified with the
pomeron.} collision as Bj\"orken-$x$ in the $\gamma p$ interaction.

\begin{floatingfigure}{0.6\textwidth}
\begin{center}
\psfig{file=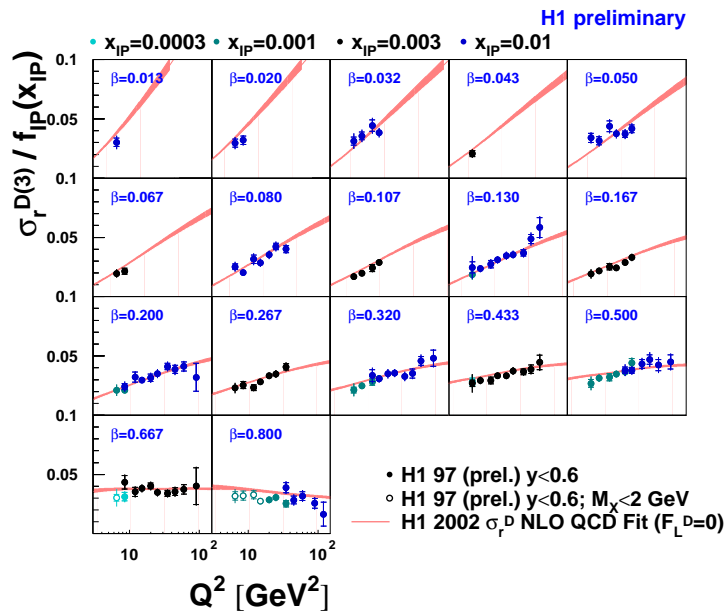,width=0.6\textwidth}
\end{center}
\caption{The diffractive cross section divided by the pomeron
flux as a function of $Q^2$ in bins of $\beta$ and for different
$\xpom$ values. The band represents the result of the NLO QCD fit
discussed in Sec.~\ref{sec:fits}.}
\label{fig:h1sigma}
\vspace{-20mm}
\end{floatingfigure}

Figure~\ref{fig:h1sigma} shows recent measu\-re\-ments obtained by the H1
Collaboration~\cite{bib:h1sigma}.  The results are presented as a
reduced cross section, $\sigma_r^{D(3)}$, defined through
\begin{multline}
\frac{d^3 \sigma^D}{d\xpom dx dQ^2} = \\
\frac{4\pi\alpha^2}{xQ^4}\left(1-y+\frac{y^2}{2}\right) \\
\times \sigma_r^{D(3)}(\xpom, x, Q^2),
\end{multline}
and divided by the ``pomeron flux'' $f_{\pom}(\xpom)$.  This flux
factor is obtai\-ned from a pa\-ra\-me\-ter\-i\-za\-tion of the $\xpom$
dependen\-ce of the cross section inspired by Regge theory.  The
parameterization works well, as can be seen from the overlap of data
points at the same $\beta$ and $Q^2$ values obtained for different
proton momentum losses $\xpom$.  The intercept of the po\-me\-ron
trajectory extracted from this data is 
\begin{equation} 
\alpha_{\pom} = 1.173 \pm 0.018 {\rm\ (stat.)} \pm
0.017 {\rm\ (syst.)} ^{+0.063}_{-0.035} {\rm\ (model)}. 
\end{equation}
Positive scaling violations are observed in most of the phase space,
suggesting a large gluon content of the diffractive exchange. The ratio
of diffractive to inclusive DIS cross sections is found to be
reasonably flat at fixed $x$ as function of $Q^2$, indicating that the same scaling
violations occur in both proccesses.

The ZEUS Collaboration has recently installed a new ``Forward Plug''
calorimeter which covers the range in pseudorapidity $4 < \eta < 5$ and
increases the measurable range in mass of the photon dissociation
system, $M_X$, to 35 GeV, while reducing the range in mass of the
proton disscociation system, $M_Y$ to 2.3 GeV and hence reducing the
background due to proton dissociation.  Preliminary
results can be found in~\cite{bib:zeussigma}.

\newpage 

\subsection{Next-to-leading order DGLAP fits}
\label{sec:fits}

\begin{floatingfigure}{0.6\textwidth}
\begin{center}
\psfig{file=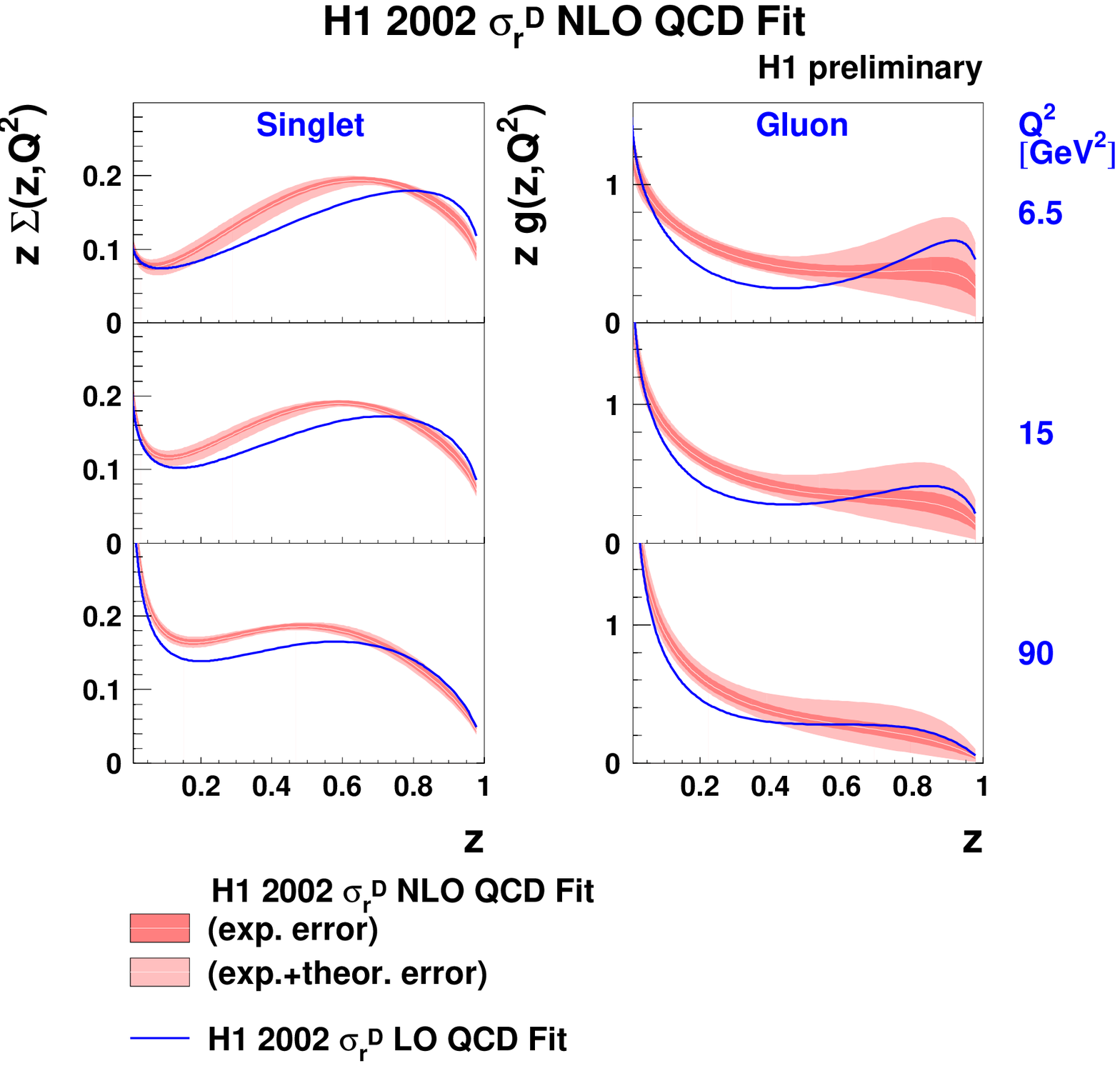,width=0.6\textwidth}
\end{center}

\caption{Diffractive quark singlet ($6*u$ with
$u=d=s=\overline{u}=\overline{d}=\overline{s}$) and gluon density.  The
pomeron flux is normalized to unity at $\xpom = 0.003$.}

\label{fig:h1fits}
\end{floatingfigure}

\noindent Using a parametrization based on Chebychev polynomials at a
starting scale of $Q^2_0 = 3 {\rm\ GeV}^2$, quark and gluon densities
have been fitted to the observed H1 cross section by applying the DGLAP
evolution equations.  Subleading reggeon exchan\-ges are included
assuming the structure function of the pion. The fit, which includes
the data shown in Fig.~\ref{fig:h1sigma} together with data at higher
$Q^2$ ($200 < Q^2 < 800 {\rm\ GeV}^2$)~\cite{bib:h1sigma_highq2} yields
a $\chi^2/ndf = 308.7/306$.  

Figure~\ref{fig:h1fits} shows the result of the next-to-leading order
(NLO) QCD fit, with full propagation of statistical, systematic
experimental and theoretical errors. The momentum fraction carried by
gluons is estima\-ted to be $75 \pm 15\% 
$. The resolved pomeron model used in the
subsequent sections uses LO parton densities (also shown in
Fig.~\ref{fig:h1fits}) and LO parton cross sections.  NLO effects are
then simulated with parton cas\-ca\-des.

\section{Factorization tests}

\begin{floatingfigure}{0.4\textwidth}
\begin{center}
\psfig{file=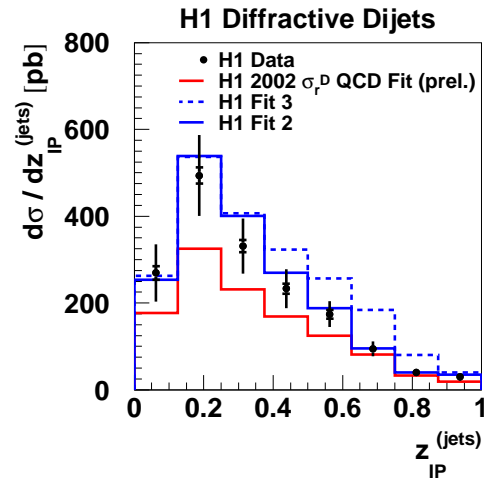,width=0.4\textwidth}
\end{center}

\caption{Diffractive DIS dijet cross section compared with the latest
LO QCD fit as well as fits to previous data.}

\label{fig:h1disdijets}
\end{floatingfigure}

The conditional parton densities obtained from the analysis of the
inclusive diffractive DIS cross section have been used succesfully to
describe the semi-in\-clu\-si\-ve cross section for dijet and open charm
production in diffractive DIS.  Both these processes are driven by the
large gluon density through boson-gluon fusion.

Figure~\ref{fig:h1disdijets} shows the cross section for diffractive
dijet production in DIS obtained by the H1
Collaboration~\cite{bib:h1jets} as a function of the fractional
momentum of the colourless exchange entering the dijet system,
$\zpom$.  The data are compared to the resolved pomeron model. 
Although the fit based on the latest H1 data yields a smaller gluon
content and therefore a lower dijet cross section than previous
fits~\cite{bib:h1sigma94}, the data are reasonably described given the
large uncertainties on the extracted gluon density as well as on the
scale for final state predictions, and missing higher
order corrections.

The $D^{\ast\pm}$ DIS production cross section as measured by
ZEUS~\cite{bib:zeuscharm} is also well described by the resolved
po\-me\-ron model.  Again, a lower cross section is observed with the model
based on the latest H1 fit as compared to previous fits.  Within
uncertainties, however, no evidence for breakdown of QCD hard
scattering factorization is observed between the inclusive and
semi-inclusive cross sections for diffractive DIS.

\begin{floatingfigure}{0.4\textwidth}

\begin{center}
\psfig{file=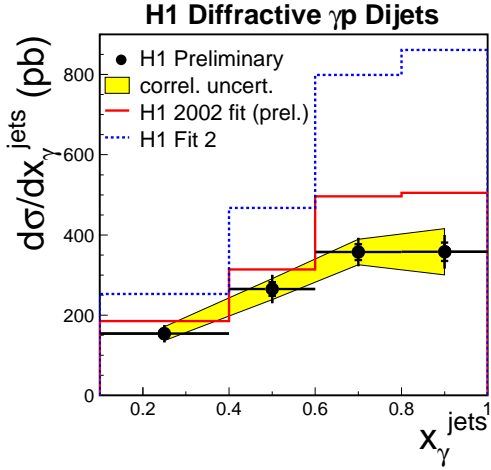,width=0.4\textwidth}
\end{center}

\caption{The dijet diffractive photoproduction cross
section as function of $x_\gamma^{jets}$. The GRV LO parton
distributions are used for the photon.}

\label{fig:h1gammapdijets}
\end{floatingfigure}

The H1 Collaboration has studied dijets in diffractive
photoproduction~\cite{bib:h1gammapjets}.  Resolved and direct processes are distinguished by
reconstructing the variable $x_\gamma$, defined as the fractional
momentum of the quasi-real photon entering the dijet system.  For
direct processes $x_\gamma$ should be unity, while in the case of
resolved processes $x_\gamma < 1$ will hold.  The hadron level variable
is $x_\gamma^{jets}$.

Figure~\ref{fig:h1gammapdijets} displays the measured cross section for
diffractive dijet photoproduction as a function of $x_\gamma^{jets}$. 
The data are compared to predictions by the resolved pomeron model
using the new H1 fit based on 2002 data and the previous ``Fit 2''.  A
model-independent evaluation of the rapidity gap suppression factor in
diffractive photoproduction relative to diffractive DIS is obtained by
calculating the double ratio of measured data to the model prediction
in photoproduction relative to DIS, yielding $1.80 \pm 0.45$.  This
factor is found to be the same, within errors, for direct and resolved
processes.

\section{Conclusion}

High precision measurements of the diffractive cross section in
DIS have been performed by H1 and ZEUS in an increased region of phase
space.  The data support Regge factorization (provided subleading
trajectories can contribute) with a value of the pomeron intercept which
is larger than for the soft pomeron.  New NLO QCD fits are available
yielding diffractive parton densities that can be used to test QCD hard
scattering factorization.

Data on the inclusive cross section and final state (open charm and
dijet production) in diffractive DIS in  are in agreement with QCD
factorization.  A study of diffractive photoproduction of dijets finds
a suppression with respect to DIS dijets, but cannot confirm different
suppression factors for resolved and direct processes.

\section*{Acknowledgments}
I am indebted to all members of the H1 and ZEUS Collaborations who
contributed to these results by collecting and analysing the
experimental data.

\end{document}